\title{Three Degrees of Distance on Twitter}
\author{
        Jorge Fabrega \\
                School of Government\\
        Adofo Ibanez University\\
        Santiago, Chile
            \and
        Pablo Paredes\\
                School of Government\\
        Adofo Ibanez University\\
        Santiago, Chile
}
\date{First version July 30, 2012.\\
This draft \today}

\documentclass[twocolumn]{article}
\usepackage{graphicx, subfigure}
\graphicspath{c:/Users/Jorge/Desktop/Investigacion/Three Degree on Twitter/ThreeDegreeOfSeparation/}

\begin{document}
\maketitle

\begin{abstract}
Recent work has found that the propagation of behaviors and sentiments through networks extends in ranges up to 2 to 4 degrees of distance. The regularity with which the same observation is found in dissimilar phenomena has been associated with friction in the propagation process and the instability of link structure that emerges in the dynamic of social networks. We study a contagious behavior, the practice of retweeting, in a setting where neither of those restrictions is present and still found the same result.

\textbf{Keywords}: Twitter, information difussion, contagion, social networks, social distance
\end{abstract}

\section{Introduction}
Recent studies have suggested that obesity\cite{obesity}, happiness\cite{hapiness}, smoking\cite{smoking} and loneliness\cite{loneliness} among others\cite{alcohol} can be contagious in social networks with up to three degrees of distance. The pattern has been named the three degree of influence rule\cite{connected}. There are two parts in this remarkable finding. On one hand, there is a claim about causation and; on the other hand, there is an empirical observation about the social distance at which a phenomena is expected to be observed by a given individual.

An open question still remains regarding the underlying mechanism of transmission in the case of causal processes \cite{socialcontagion,unfriending,lyon}. For example, in the case of obesity, one mechanism that could explain the observation of group formation around individuals’ weight is the spread of eating habits. One could consider the case of when the friend of your friend starts eating some unhealthy food whereby the probability that your friend will gain weight increases as a by product of their friendship and, later, so does your probability. An alternative explanation may rest on individuals’ perception; for instance, an individual could ignore that he is unhealthily gaining weight because people surrounding him are also becoming fatter. Finally, the observation may simply be a correlation resulting from the human inclination to form bonds with people that look like themselves. Only well-designed experiments will be able to identify the ruling mechanism in each case\cite{shoham}. 

Nevertheless, the claim regarding social distance is a different story. Three degree of influence implies that if you pick a randomly selected individual in a network and observe her behavior, feelings or preferences, you should expect to also observe, with a given probability, similar behaviors, feeling or preferences in her relevant network at up to between 2 and 4 degrees of separation in most studied phenomena\cite{socialcontagion}. The regularity with which the same observation is found in dissimilar phenomena is, by itself, a black box. An explanation consistent with the social contagion hypothesis would be that the “technology” to produce social contagion is resource intensive (e.g. time consuming) and consequently subject to diminishing returns to scale. Based on the dynamic of links’ formation, Christakis and Fowler\cite{socialcontagion} have suggested that given that closer ties tend to be more stable than farther ones the set of individuals at greater social distances change at a faster rate reducing the capacity of the spreading mechanism to act on them.

It would be interesting to observe a social contagion phenomenon in a context without those restrictions (i.e one in which friction to spread the contagion is negligible and social structure is stable) in order to verify whether it spreads to farther distances, as would be expected when they are not binding.

To advance in that direction, we propose to examine the information diffusion of messages in the social media Twitter as an example of social contagion. In Twitter, users share opinions and information with other users who decided to follow that user’s messages (tweets). Those tweets cannot be larger than 140 characters. Other users can resend tweets (retweets) to their own followers by pressing a button. As a consequence, a tweet can travel from user to user. The practice of retweeting a tweet happens in a fundamentally stable and friction-free environment. First, the dynamic of retweets extends for short periods of time after the original message was created\cite{vanliere}; consequently, the topology of the network surrounding a given user will not change significantly during the time in which contagion can occurs, and, second, the costs to produce a retweet are negligible such that the technology of the retweets can be taken as frictionless. Therefore, given that friction and instability of the social network should not be binding restrictions to the spread of contagion, we should expect that the spread of tweets along Twitter’s social graph should reach higher social distances than those found in social phenomena were those restrictions are supposedly active.

Many messages written in Twitter are resent by other users without manipulation. However, a user who follows two or more of the users who sent the message (either the original or a retweet) receives it only once. In this way, Twitter Co.\ avoid filling their users’ timelines who follow two or more of those retweeterers with the same repeated message again and again. For example, consider a subgraph formed by four nodes (user0 to user3) where user2 and user3 are followers between them and followers of user1. And, suppose that user1 follows user0. If user1 has retweeted a message from user0 and, later, user3 do the same, user2 will only receive the tweet through the path user0$\rightarrow$user1 and not through the path user0$\rightarrow$user1$\rightarrow$user3. Consequently, the distance travelled by a given tweet along the underlying social graph connecting twitter accounts can be tracked. 

We will call the longest path between the original sender and the farthest retweeter the degree of influence on Twitter to keep in mind the connection with Christakis and Fowler, although we are aware that the measure of influence on Twitter has been a matter of controversy and it is still an open question \cite{sinanaral}. However, our focus is not the analysis of influence by itself but the measuring of the social distance reached by a frictionless social contagion process.

\section{Methodology and Data}
For access to the data, Twitter offers an Application Programming Interface (API) split in three formats: Streaming API for real-time tweets, the Search API for past tweets and the REST API for specific queries about tweets and users. The first ones does not have important restrictions in the amount of queries, however, the third one has a limit of 350 calls per hour. For this reason, we complement the REST API calls with an external proxy service called Apigee, that allowed us to continue performing queries after the depletion of our API calls on the official Twitter services. However, as a proxy service, the calls are slower than with the original API.

For the purpose of this study, we collected tweets, their retweets and the social graph connecting the accounts of tweeters (the authors of original messages) and retweeters (the re-senders of those messages) according to the follower/following relationship between the users. The methodology was the following. 

First, using Twitter Streaming API (Application Programming Interface), we obtained 3 million tweets in a real-time period of 24 hours (2012/06/01), collecting information about its sender, the time at which was created, among other information. As mentioned above, in a network, one potential reason for the declining diffusion of a behavior at farther distances is the the higher instability of social structure as distance increases from a given individual. To control for that eventuality, it was required to recover the social graph of a retweeted tweet in a short period of time. Unfortunately, Twitter imposes rate-limit to access to the REST API services used for this purpose. As a consequence, when a user has many followers, even using Apigee proxy, the recovering of the social graph is slow. Therefore, there is a risk that the observed graph had changed between the moment when the original tweet was sent and the instance in which the underlying social graph of followers and friends was generated. For this reason we opted to reduce the number of tweets for further analysis and instead of studying the 3 million tweets,  we randomly selected a subsample of 400,000 tweets. 

Second, for each tweet in the dataset, we verified its status through the Twitter REST API, recovering the numbers of retweets that it received (if any) and the retweeter's ids for each tweet. We have taken a conservative approach to the process of information diffusion, specifically, we have focused our attention on native mechanisms of retweeting, which means we will consider as a retweeted tweet any tweet that Twitter API identifies as such. We selected this operationalization to observe cases of pure contagion. Recent work \cite{azman} has suggested broader definitions for retweeting behavior and it is a matter of future research to test whether the results presented here hold in those cases. 

\begin{figure}
\begin{center}
\includegraphics[scale=0.5,width=0.5\textwidth]{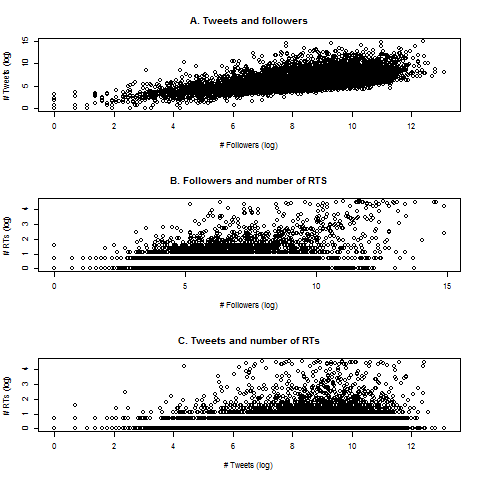}
\label{Figure 1}
\caption{Tweets, Retweets and Followers}
\end{center}
\end{figure}

The result gave us a total of 13,946 retweeted tweets (3.5\%). This proportion is consistent with those found in previous studies \cite{suh}. Approximately, 76.4\% of the tweets were retweeted only once; 12,6\% twice and 11\% three or more times. As expected, we found a clear association between tweeters’ number of tweets and their number of followers (figure 1a) and between the number of followers and amount of retweets (figure 1b); although, these data suggest that more active tweeters are not necessarily more retweeted than less active ones (figure 2c). 

Third, for each retweeted tweet, we rebuilt the social graph of friends and followers linking the tweeter with each retweeter using Twitter REST API. Therefore, we generated 13,946 social subgraphs, one per retweeted tweets. Then, for each social graph, we compute the eccentricity of the tweeter’s tweet (i.e the longest geodesic connecting each tweeterer with the set of the retweeters of her tweet). As shown in figure 2 to 5, there were variation in eccentricities and number of retweets.

\begin{figure}[ht!]
	\begin{center}
        	\subfigure{}
            	\label{first}
		\caption{1 degree 95 RTs}
            	\includegraphics[scale=0.25,width=0.3\textwidth]{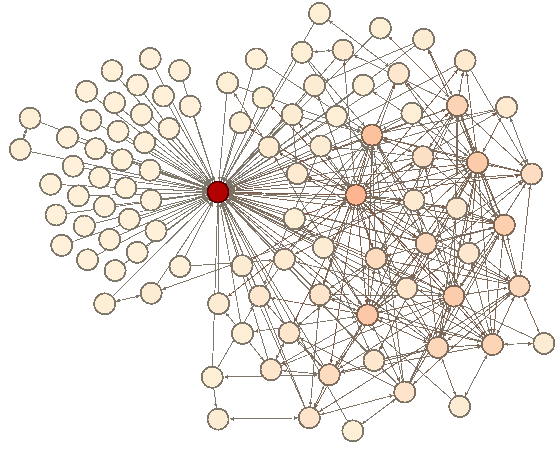}
        	\subfigure{}
           	\label{second}
		\caption{1 degree 3 RTs}
           	\includegraphics[scale=0.25,width=0.2\textwidth]{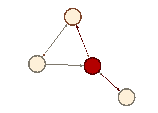}
	\end{center}
\end{figure}

\begin{figure}[ht!]
	\begin{center}
        	\subfigure{}
            	\label{third}
		\caption{2 degrees 49 RTs}
            	\includegraphics[scale=0.25,width=0.25\textwidth]{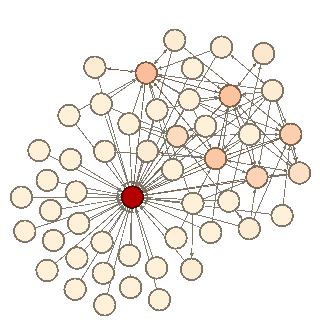}
        	\subfigure{}
            	\label{fourth}
		\caption{2 degrees 3 RTs}
            	\includegraphics[scale=0.25,width=0.3\textwidth]{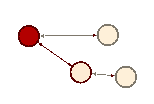}
	\end{center}
\end{figure}

\section{Results}\label{results}
Remarkably, in a network of followers, the social distances travelled by retweeted tweets are in the same range found for other phenomena in the literature of social contagion. Nevertheless, friction in the propagating mechanism and instability of ties should not be conditioning the spread of contagion in this case. We found that 87\% of retweeted tweets (12,126 tweets) were retweeted at one degree of distance, 7\% (965 tweets) were retweeted up to two degrees of distance, 1\% (156 tweets) travelled up to the third degree of distance and 0.5\% of retweeted tweets were resent to farther distances including one tweet that traveled up to the ninth degree of distance via retweets. In the remaining 4.6\% of cases, there is a mixed pattern where some retweets were made by unconnected accounts (figure 6). Hence, even in cases with a significant number of retweets, the audience remain fundamentally local from a social structural perspective.

\begin{figure}
\begin{center}
\includegraphics[trim=10mm 10mm 10mm 10mm, scale=0.5]{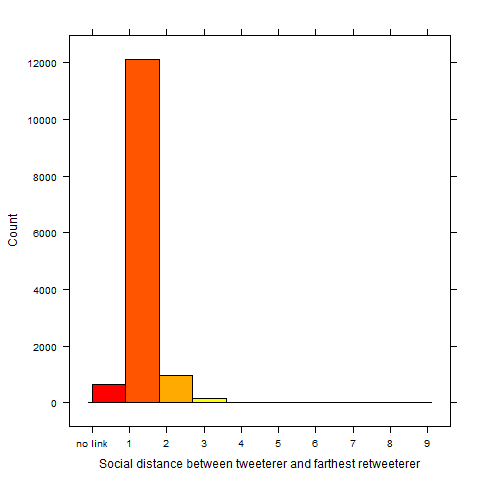}
\label{Figure 6}
\caption{Eccentricities of retweeted tweets}
\end{center}
\end{figure}

On the other hand, as shown in figure 7, there is a positive although weak relationship between the number of retweets and the social distance travelled by a tweet. We also analyzed smaller samples of tweets and performed basic statistical analyses to verify our reading of the data and found the same results. 

\begin{figure}
\begin{center}
\includegraphics[trim=10mm 10mm 10mm 10mm, scale=0.5]{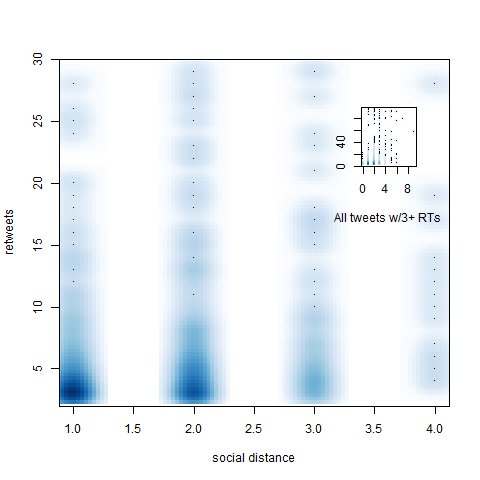}
\label{Figure 7}
\caption{Relationship between retweets and social distance with 3+ RTs}
\end{center}
\end{figure}

\section{Discussion}\label{discussion}
Recent work has found that the propagation of behaviors and sentiments through networks extends in ranges up to 2 to 4 degrees of distance. This finding was so remarkable that Christakis and Fowler\cite{connected} decided to give it a name: the three degree of influence rule. From the perspective of a theory of contagion, propagation does not extend to greater distances because either the mechanism of diffusion becomes weaker or the dynamic of link formation and destruction negatively affects the reachability of individuals located at greater distances.. We have proposed to measure the propagation of a behavior along a social network in circumstances in which neither of these constraints should be binding and found, contrary to  our expectation, that the diffusion remains in the same range found in the literature.

One alternative explanation, based on homophily, would suggest that the decline at further distances is simply because dissimilar individuals are located at larger social distances from each other. Such a possibility would be consistent with a practice of retweeting as an expression of homophily. However, recent work\cite{retweet} offers evidence in the opposite direction suggesting greater levels of anti-homophily in retweeting behaviors. Further research is required to explain why social contagion usually stops up to the third degree of distance.

\bibliographystyle{abbrv}

\end{document}